\documentclass[aps,prl,twocolumn,superscriptaddress]{revtex4-1}

\usepackage{amsmath}
\usepackage{graphicx}
\usepackage{float}
\usepackage{multirow}
\usepackage{lineno}
\usepackage{color}
\usepackage[normalem]{ulem}
\usepackage{soul}

\newcommand{\hsix}{$\rm ^{6}H~$}
\newcommand{\li}{$\rm ^{7}Li$}
\newcommand{\pip}{$\pi^{+}$}

\setstcolor{red}

\begin{document}

\title{Measurement of \hsix ground state energy in an electron scattering experiment at MAMI-A1}

\author{Tianhao Shao}\email{thshao21@m.fudan.edu.cn}
\affiliation{Institut für Kernphysik, Johannes Gutenburg-Universität, D-55099, Mainz, Germany}\affiliation{Key Laboratory of Nuclear Physics and Ion-beam Application (MOE), Institute of Modern Physics, Fudan University, 200433, Shanghai, China}

\author{Jinhui Chen}\email{chenjinhui@fudan.edu.cn}
\affiliation{Key Laboratory of Nuclear Physics and Ion-beam Application (MOE), Institute of Modern Physics, Fudan University, 200433, Shanghai, China}\affiliation{Shanghai Research Center for Theoretical Nuclear Physics, NSFC and Fudan University, 200433, Shanghai, China}

\author{Josef Pochodzalla}\email{pochodza@uni-mainz.de}
\affiliation{Institut für Kernphysik, Johannes Gutenburg-Universität, D-55099, Mainz, Germany}
\affiliation{Helmholtz-Institut Mainz, Johannes Gutenberg-Universit\"at Mainz, 55099 Mainz, Germany}
\affiliation{PRISMA$^+$ Cluster of Excellence, Johannes Gutenberg-Universit\"at Mainz, 55099 Mainz, Germany}

\author{Patrick Achenbach}\affiliation{Institut für Kernphysik, Johannes Gutenburg-Universität, D-55099, Mainz, Germany}
\author{Mirco Christmann}\affiliation{Institut für Kernphysik, Johannes Gutenburg-Universität, D-55099, Mainz, Germany}
\author{Michael O. Distler}\affiliation{Institut für Kernphysik, Johannes Gutenburg-Universität, D-55099, Mainz, Germany}
\author{Luca Doria}\affiliation{Institut für Kernphysik, Johannes Gutenburg-Universität, D-55099, Mainz, Germany}
\author{Anselm Esser}\affiliation{Institut für Kernphysik, Johannes Gutenburg-Universität, D-55099, Mainz, Germany}
\author{Julian Geratz}\affiliation{Institut für Kernphysik, Johannes Gutenburg-Universität, D-55099, Mainz, Germany}
\author{Christian Helmel}\affiliation{Institut für Kernphysik, Johannes Gutenburg-Universität, D-55099, Mainz, Germany}
\author{Matthias Hoek}\affiliation{Institut für Kernphysik, Johannes Gutenburg-Universität, D-55099, Mainz, Germany}
\author{Ryoko Kino}\affiliation{Graduate School of Science, Tohoku University, Sendai, Miyagi 980-8578, Japan}
\affiliation{Graduate Program on Physics for the Universe (GP-PU), Tohoku University, Sendai, Miyagi, 980-8578, Japan}
\author{Pascal Klag}\affiliation{Institut für Kernphysik, Johannes Gutenburg-Universität, D-55099, Mainz, Germany}
\author{Yu-Gang Ma}\affiliation{Key Laboratory of Nuclear Physics and Ion-beam Application (MOE), Institute of Modern Physics, Fudan University, 200433, Shanghai, China}\affiliation{Shanghai Research Center for Theoretical Nuclear Physics, NSFC and Fudan University, 200433, Shanghai, China}
\author{David Markus}\affiliation{Institut für Kernphysik, Johannes Gutenburg-Universität, D-55099, Mainz, Germany}
\author{Harald Merkel}\affiliation{Institut für Kernphysik, Johannes Gutenburg-Universität, D-55099, Mainz, Germany}
\author{Miha Mihovilovi\v{c}}\affiliation{Institut für Kernphysik, Johannes Gutenburg-Universität, D-55099, Mainz, Germany}
\author{Ulrich M\"uller}\affiliation{Institut für Kernphysik, Johannes Gutenburg-Universität, D-55099, Mainz, Germany}
\author{Sho Nagao}\affiliation{Department of Physics, Graduate School of Science, The University of Tokyo, 113-0033, Tokyo, Japan}
\author{Satoshi N. Nakamura}\affiliation{Graduate School of Science, Tohoku University, Miyagi 980-8578, Sendai, Japan}
\author{Kotaro Nishi}\affiliation{Department of Physics, Graduate School of Science, The University of Tokyo, 113-0033, Tokyo, Japan}
\author{Ken Nishida}\affiliation{Department of Physics, Graduate School of Science, The University of Tokyo, 113-0033, Tokyo, Japan}
\author{Fumiya Oura}\affiliation{Graduate School of Science, Tohoku University, Sendai, Miyagi 980-8578, Japan}
\affiliation{Graduate Program on Physics for the Universe (GP-PU), Tohoku University, Sendai, Miyagi, 980-8578, Japan}
\author{Jonas Pätschke}\affiliation{Institut für Kernphysik, Johannes Gutenburg-Universität, D-55099, Mainz, Germany}
\author{Bj\"orn S\"oren Schlimme}\affiliation{Institut für Kernphysik, Johannes Gutenburg-Universität, D-55099, Mainz, Germany}
\author{Concettina Sfienti}\affiliation{Institut für Kernphysik, Johannes Gutenburg-Universität, D-55099, Mainz, Germany}
\author{Daniel Steger}\affiliation{Institut für Kernphysik, Johannes Gutenburg-Universität, D-55099, Mainz, Germany}
\author{Marcell Steinen}\affiliation{Helmholtz-Institut Mainz, Johannes Gutenberg-Universit\"at Mainz, 55099 Mainz, Germany}
\author{Michaela Thiel}\affiliation{Institut für Kernphysik, Johannes Gutenburg-Universität, D-55099, Mainz, Germany}
\author{Andrzej Wilczek}\affiliation{Institute of Physics, University of Silesia in Katowice, 41-500, Chorzów, Poland}
\author{Luca Wilhelm}\affiliation{Institut für Kernphysik, Johannes Gutenburg-Universität, D-55099, Mainz, Germany}

\collaboration{A1 Collaboration}

\begin{abstract}
For the first time the neutron-rich hydrogen isotope \hsix was produced in an electron scattering experiment in the reaction $\rm ^{7}Li(e,~e'p\pi^{+})^{6}H$ using the spectrometer facility of the A1 Collaboration at the Mainz Microtron accelerator. By measuring the triple coincidence between the scattered electron, the produced proton, and $\pi^{+}$, the missing mass spectrum of \hsix was obtained. A clear peak above $^3$H+n+n+n energy threshold was seen resulting in a ground state energy of \hsix at $2.3\pm0.5({\rm stat.})\pm0.4({\rm syst.})$~MeV with a width of $1.9\pm1.0({\rm stat.})\pm0.4({\rm syst.})$~MeV. This work challenges the understandings of multi-nucleon interactions and presents a new method to study light neutron-rich nuclei with electron scattering experiments.
\end{abstract}

\maketitle

$Introduction-$One of the most basic questions in nuclear physics is the maximum number of the neutron that can be bound in a nucleus with a given number of proton. Experimental efforts to search for the nuclei far beyond from the neutron drip line, such as tetraneutron \cite{Duer:2022ehf} and $\rm ^{28}O$ \cite{Kondo:2023lty}, have been carried out recently. For the fundamental isotope hydrogen containing only one proton, besides its well-known isotopes deuteron and triton, several very neutron-rich isotopes from $\rm ^{4}H$ to $\rm ^{7}H$ have been observed \cite{Belozyorov:1986skf,Caamano:2008zz,Korsheninnikov:2001buf,Aleksandrov:1984tb,Korsheninnikov:2003bz,Gurov:2009,Bezbakh:2019dvh}. Especially \hsix and $\rm ^{7}H$, who present the largest neutron-to-proton ratios known so far, are unique platforms to study the NN and multi-N interactions in a neutron-rich environment. However, knowledge of them are still not well established both on the experimental and theoretical sides.

\hsix was first observed in the reaction $\rm ^{7}Li(^{7}Li,~^{8}B)^{6}H$ in 1984 \cite{Aleksandrov:1984tb}. With respect to the structure of $\rm ^{3}H$+n+n+n, an unbound \hsix ground state was found at an energy of resonance $E_{r} = 2.7 \pm 0.4$~MeV and a width $\Gamma = 1.8 \pm 0.5$~MeV. This result was confirmed with $E_{r} = 2.6 \pm 0.5$~MeV in the reaction $\rm ^{9}Be(^{11}B,~^{14}O)^{6}H$ \cite{Belozyorov:1986skf}. However these two observations could not be reproduced in the latter experiments with pion beams. In the reaction $\rm ^{9}Be(\pi^{-},~pd)^{6}H$ a spectrum of states with the lowest energy at $E_{r} = $~6.6~$\pm$~0.7~MeV and a width $\Gamma = $~5.5~$\pm$~2.0~MeV were found in 2003 \cite{Gurov:2003pv} in which no state was seen below 3~MeV. Similar conclusion could also be drawn in the reaction $\rm ^{11}B(\pi^{-},~p^{4}He)^{6}H$ with a ground state (g.s.) energy at $E_{r} = $~7.3~$\pm$~1.0~MeV. About 5 counts of \hsix were seen in the reaction $\rm ^{12}C(^{8}He,~^{14}N)^{6}H$ in 2008 \cite{Caamano:2008zz}. The result of the g.s.~energy at $E_{r} = $~2.91$^{+0.85}_{-0.95}$~MeV with a width $\Gamma = $~1.52$^{+1.77}_{-0.35}$~MeV favored the 1984 result. The most recent experimental study on $\rm ^{6}H$ used the reaction $\rm ^{2}H(^{8}He,~^{4}He)^{6}H$~\cite{Nikolskii:2021kqe}. Due to the indistinct spectrum structure, a lower limit for the g.s.~energy was given at about 4.5~MeV with a large width favoring the larger \hsix energy. Several negative experimental studies without any signal of \hsix are also reported \cite{Parker:1990sla}. Different experimental techniques report large differences on \hsix g.s.~energy and width. More independent measurements on \hsix energy with different reactions are expected to establish the knowledge of it. 

The understanding of \hsix is also ambiguous in theoretical calculations. After the first observation of $\rm ^{6}H$, its g.s.~energy was calculated to be at about 5.5~MeV \cite{Poppelier:1985vpj} and 1.3~MeV \cite{Bevelacqua:1986zz} within the nuclear shell model, respectively. By using the method of angular potential functions, an energy at 6.3~MeV was reported in 1989 \cite{Gorbatov:1989}. The antisymmetrized molecular dynamics model has also been applied in calculation of \hsix and an energy of 6.6~MeV was obtained in 2004 \cite{Aoyama:2004tay}. Recently a Gamow shell model calculation gave an energy of about 3.2~MeV with a width of about 2~MeV \cite{Li:2021tyy} favoring the \hsix observations with smaller g.s.~energies. However the most recent calculation published in 2022, in which the energies of $\rm ^{4}H$ and $\rm ^{5}H$ were reproduced with an effective n-$\rm ^{3}H$ potential, predicted a large \hsix g.s.~energy at about 10~MeV with a large width of about 4~MeV \cite{Hiyama:2022gzv}.

In conclusion the research on \hsix should be continued both experimentally and theoretically. In this Letter \hsix is observed for the first time in an electron scattering experiment using the reaction $\rm ^{7}Li(e,~e'p\pi^{+})^{6}H$. By reconstructing the missing-mass spectrum of $\rm ^{6}H$, the g.s.~energy and corresponding width of it are measured.

$Experiment-$The experiment was performed with the spectrometer facility of the A1 Collaboration~\cite{Blomqvist:1998xn} at Mainz Microtron (MAMI). A 0.75~mm wide and 45~mm long natural lithium target (2.4~g/cm$^2$ mass thickness), composed with about 92.7\% $\rm ^{7}Li$ and 7.3\% $\rm ^{6}Li$, was bombarded with a 855~MeV electron beam along its long direction with beam current of 400~nA (more details in ref. \cite{Eckert:2022srr}). When the energy transferred from the beam electron to a proton in \li~nucleus via a virtual photon is larger than 300~MeV, the proton can be excited into a $\Delta^+(1232)$ which is short-lived and decays to a neutron and a \pip. The neutron from decay carrying residual momentum can scatter off another proton in the nucleus. In the condition where the proton absorbs the momentum of the neutron and leaves the nucleus, a \hsix nucleus is produced. With momenta of the scattered electron, the produced proton and the \pip, the energy spectrum of \hsix can be obtained by reconstructing the missing mass in the reaction by:

\begin{widetext}
\begin{align}
    \label{eq1}
    E_{r}~=~&\sqrt{
                     (E_{\rm e}+M_{\rm ^{7}Li}-E_{e'}-E_{\rm p}-E_{\pi^{+}})^2
                     -(\mathbf{p}_{\rm e}-\mathbf{p}_{\rm e'}-\mathbf{p}_{\rm p}-\mathbf{p}_{\rm \pi^+})^2} 
                ~-~M_{\rm {^3}H+3n}.
\end{align}
\end{widetext}

\begin{figure}[!htbp]
  \centering
    \includegraphics[width=0.4\textwidth]{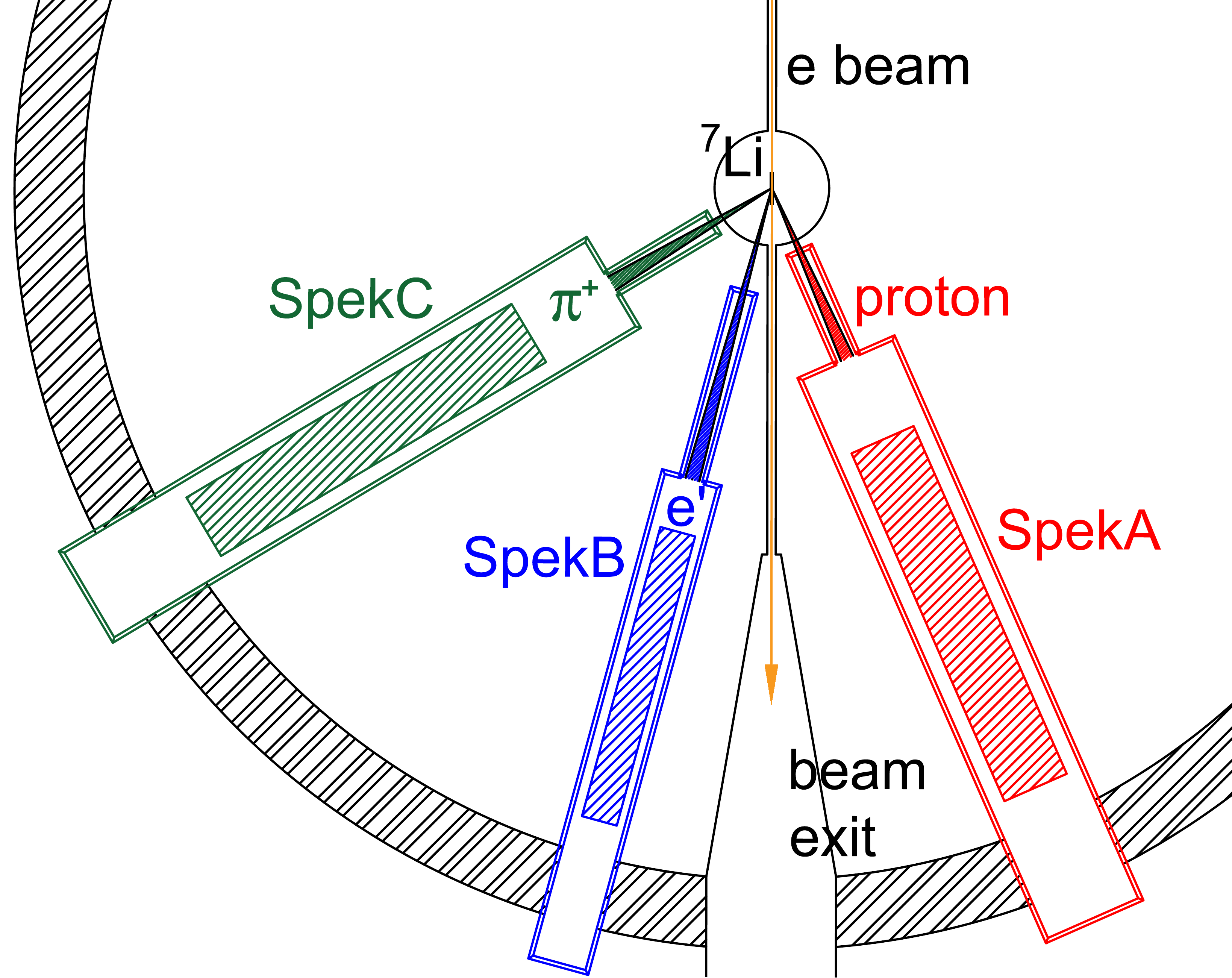}
    \caption{Illustration of the experiment setup in A1 hall. The red, blue, and green spectrometers are SpekA, B, and C, respectively. The beam enters from the top. Shaded squares illustrate the detector regions. The angular acceptances of them are 28~msr (A), 5.6~msr (B), and 28~msr (C).}\label{fig1}
\end{figure}

Three high resolution spectrometers SpekA, SpekB, and SpekC mounted on the annular track around the target are used to measure the momentum of produced proton, scattered electron, and produced \pip, respectively. According to the kinematics of the reaction, SpekA, B, and C are set to the angles -23.8$^\circ$, 15.1$^\circ$, and 59.1$^\circ$ with respect to the beam direction as illustrated in Fig.~\ref{fig1}. By setting the magnetic fields of spectrometers, the central momenta of them are set to 417~MeV/$c$ (A), 421~MeV/$c$ (B), and 273~MeV/$c$ (C) to filter the particles with designated charge and within corresponding momenta acceptances of 20\%, 15\%, and 25\%, respectively. The directions and values of momenta are measured by vertical drift chambers (VDCs) whose resolutions can reach about 3 mrad for the angles and about 0.1\% for the values of momenta. Two layers of plastic scintillation counters named dE and Time of Flight (TOF) are used to measure the energy losses and coincidence timings of particles. Electrons are discriminated from the pions by the gas Cherenkov detector. The momenta and vertexes of particles at target position are obtained by transporting the measured raw momenta from VDCs to the target position according to their trajectories in magnetic fields of spectrometers. The reaction vertex is determined by SpekB since it has the best vertex resolution at about 1~mm on longitudinal direction due to its specific magnet configuration~\cite{Blomqvist:1998xn}.

The production times of particles, which are used to make timing coincidences, are obtained by subtracting the times of flight in their trajectories, which are from target position to TOF positions, from the measured times by TOF. During the data acquisition the double timing coincidences of each spectrometer pairs (AB, AC, and BC) were required.

With a Monte Carlo (MC) simulation taking into account the resolutions of three spectrometers and the energy losses of particles in their trajectories from the target to the detectors, the evaluated resolution of \hsix missing-mass spectrum is about 1.2~MeV. Taking into account the cross sections for the electron production of a $\Delta^+(1232)$, the n-p elastic scattering occurring within the target nucleus, and the probability of triple coincidence being recorded by three spectrometers, the expected production rate of \hsix is about 1 count per day in the interested missing mass region at 0~-~10~MeV/$c^2$.

The 7.3\% $\rm ^{6}Li$ nucleus existing in the natural lithium target may change the reaction to $\rm ^{6}Li(e,~e'p\pi^{+})^{5}H$. To evaluate the influence of it on the missing-mass spectrum, a beam time with an enriched \li~target, in which the purity of \li~is larger than 99.99\%, has also been carried out.  No obvious difference on the missing-mass spectrum is seen.

$Data~analysis-$The data taking was performed during the years 2023 and 2024 with a total beam time of about 550 hours. The produced proton, scattered electron, and produced \pip~are identified according to their specific energy losses in scintillators dE and TOF of SpekA, B, and C, respectively. To select three particles from the same reaction event, the triple coincidence between three spectrometers are required (see appendix for details). 

The missing-mass spectrum is calculated by Eq.~\ref{eq1} with the beam energy, the mass of target nucleus $M_{\rm ^{7}Li}=6533.834~{\rm MeV/}c^{2}$~\cite{Kondev:2021lzi}, and the measured momenta of scattered electron, produced proton, and \pip. The beam energy is corrected by the energy loss of beam electron in lithium target according to the measured position of reaction vertex along the beam direction. The momenta of particles are also corrected by taking into account their energy losses in the target, a few cm of air between target chamber and spectrometers, and two Kapton window foils with 125~$\mu$m thickness each, which are calculated by Bethe-Bloch formula. The energy loss corrections on the momenta are +0.16\%, +0.04\%, and +0.05\% in average for proton, scattered electron, and \pip, respectively.

Calibrations of momentum measurements are done by measuring the momenta of elastic-scattered electrons with a $\rm ^{181}Ta$ or a $\rm ^{12}C$ target after the \hsix beam time. The central momenta and angles of three spectrometers for the calibration runs are set to be similar to the \hsix experiment. The expected momenta of the scattered electron are determined by the selected beam energies and scattering angles.

For the measured momenta, the energy loss corrections were done by the same method as the \hsix experiment. Comparing these measured momenta with the expected ones, additional momentum corrections are determined to be +0.032\%, +0.087\%, and -0.119\% for the momenta measured by SpekA, B, and C, respectively. The precisions of these calibrations are $\delta p_{\rm calib}=\pm10$~keV/$c$, which were also seen in previous experiments at A1~\cite{A1:2015isi,A1:2016nfu,A1:2021sku,A1:2022wzx} and are considered as one of the systematic uncertainties.

\begin{figure}[!htbp]
  \centering
    \includegraphics[width=0.45\textwidth]{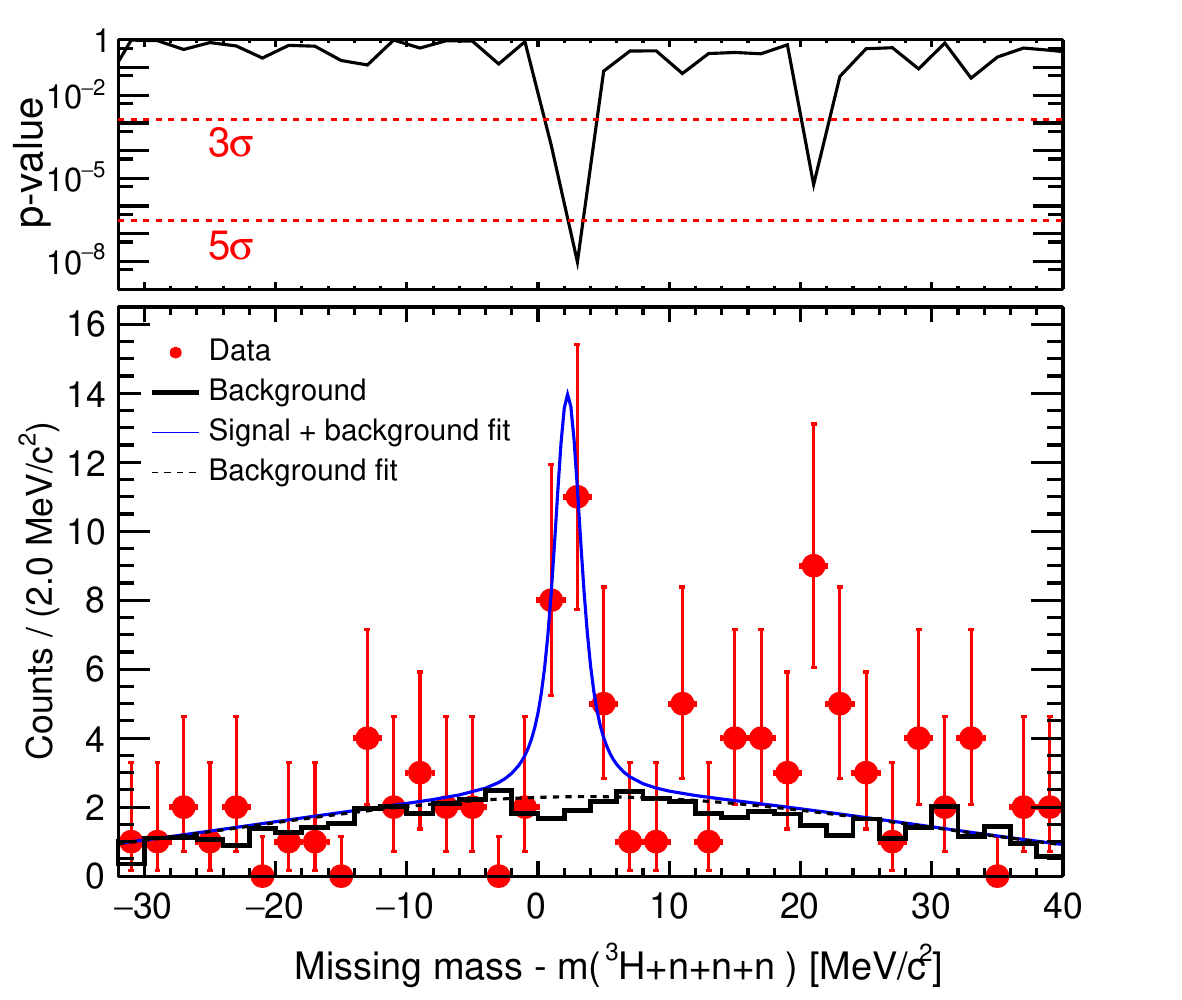}
    \caption{Bottom panel: the spectrum of reconstructed missing mass minus the \hsix threshold $\rm ^{3}H+n+n+n$. Data points are shown with red points and the black histogram represents the background from random coincidence. Error bars represent statistical uncertainties from poisson distributions. Curves are signal and background fits as described in the text. Upper panel: p values of data points in every bin with respect to the background distribution. The red dash lines represent the positions of 3$\sigma$ and 5$\sigma$ of significance.}\label{fig3}
\end{figure}

Figure~\ref{fig3} shows the spectrum of the reconstructed missing mass minus the mass of \hsix threshold $\rm ^{3}H+n+n+n$. A clear peak shows up at the energy of 2-3~MeV, which comes from the ground state of $\rm ^{6}H$. The total count of this peak is consistent with our estimated \hsix production rate. The black histogram representing the evaluated background from random coincidence describes the background shape well, especially in the region on the left of the peak. According to the MC simulation, a possible signal of the $^5$H ground state produced by the $^6$Li content is expected to appear at about -6~MeV/$c^2$ on the missing-mass spectrum of $^6$H with the order of 1 count during the measuring time. In that region the observed background is fully described by the random background, indicating that the $^5$H contamination is negligible. 

On the right side of the peak the background shape is not as good as the left side, which may be caused by some broad \hsix excited states. A peak-like structure may arise at an energy around 21~MeV. It could be related to the highest \hsix excited state at an energy of 21-22~MeV also seen in the experiments with pion beams \cite{Gurov:2003pv,Gurov:2009}. However the statistics of them are not sufficient to draw any solid conclusion on \hsix excited states. By using the RooFit~\cite{Verkerke:2003ir} an unbinned fitting with a function containing a signal part, which is formed by a Breit-Wigner distribution for signal shape convoluted with a Gaussian function whose FWHM is fixed by the missing-mass resolution of 1.2~MeV obtained from MC simulation, and a wide Gaussian distribution for background shape is applied to fit the spectrum. The obtained mean and width of the Breit-Wigner distribution by fitting are the measured energy $E_{r}$ and width $\Gamma$ of $\rm ^{6}H$. The significance of the signal calculated by the asymptotic formula \cite{Cowan:2010js} is $N_{\sigma}=\sqrt{2[(N_{\rm s}+N_{\rm b})\ln{(1+N_{\rm s}/N_{\rm b})}-N_{\rm s}]}=5.2$, in which $N_{\rm s}$ and $N_{\rm b}$ are the counts of signal and background in signal region 0~-~6~MeV/$c^2$.

According to Eq.~\ref{eq1}, the systematic uncertainties of the energy mainly come from the uncertainties of the beam energy and the momentum measurements of the three particles. The absolute accuracy of the beam energy measured by the accelerator system is $\delta E_{\rm beam}=\pm 0.16$~MeV. This uncertainty is also translated into momentum measurements while doing momentum calibrations. Uncertainties coming from setups of spectrometer angles are about $\pm$0.001~MeV/$c$ and negligible. To monitor the stability of the central momentum during the whole beam time, the magnet fields of spectrometers were measured precisely every one hour. They varied in a range of $\pm$0.04~MeV/$c$. To check the linearity of the momentum in the acceptance range, the central momenta of the spectrometers were changed in steps by tuning the magnetic field to scan the whole acceptance range during the calibration beam time. The linearity shifts of momentum measurements are found to be $\pm 0.05$~MeV/$c$. Due to the use of a long lithium target, the momentum measurements at different vertexes along the beam direction are also checked and uncertainties of $\pm 0.02$~MeV/$c$ are found. Due to the width of the $^7$Li target is 0.75~mm which is smaller than the vertex resolution, the vertex on vertical direction is always fixed on the target center. This causes $\pm$0.1~MeV/$c$, $\pm$0.02~MeV/$c$, and $\pm$0.02~MeV/$c$ uncertainties on the energy loss corrections for momenta of SpekA, B, and C, respectively. The combined systematic uncertainties for momentum measurements of SpekA, B, and C are $\delta p_{\rm A}=\pm 0.2$~MeV/$c$, $\delta p_{\rm B}=\pm 0.17$~MeV/$c$, and $\delta p_{\rm C}=\pm 0.16$~MeV/$c$, respectively. By using the bootstrap method \cite{Efron:1979bxm}, the systematic uncertainties of beam energy and momentum measurements are translated into the calculated energy according to Eq.~\ref{eq1}, which is $\pm$0.3~MeV. On the longitudinal direction of reaction vertex which is calibrated with a 6~$\mu$m thin $\rm ^{181}Ta$ target, due to the resolution and tilt of the target ladder, $\pm 2.5$~mm shift on longitudinal vertex is added which causes $\pm 0.2$~MeV uncertainty of the measured energy and $\pm 0.05$~MeV uncertainty of its width. The fixed vertical vertex causes $\pm 0.04$~MeV uncertainty of the measured energy according to the MC simulation. Systematic uncertainty due to the fitting procedure mainly comes from the background shape. Different fit regions and parametrization of the background are tested and $\pm 0.05$~MeV uncertainty for the mean value and $\pm 0.3$~MeV uncertainty for the width are found. The missing-mass resolution is changed from 1.0~MeV to 1.5~MeV, which comes from the uncertainty of the MC simulation, and $\pm 0.2$~MeV uncertainty is found for the width. The systematic uncertainties for the energy and width of g.s. \hsix are summarized in Table~\ref{tab1} and Table~\ref{tab2}, respectively. Finally the total values are 0.4~MeV for the energy and 0.4~MeV for the width.
\begin{table}[!htbp]
    \centering
    \caption{Systematic uncertainties for the g.s. energy of $^6$H.}
    \label{tab1}
    \setlength{\tabcolsep}{4mm}{
    \begin{tabular}{ccc}
    \hline
    \hline
    Uncertainty source   &Uncertainty (MeV) \\
    \hline
    Momentum measurements         &0.3  \\
    Longitudinal vertex          &0.2  \\
    Vertical vertex                 &0.04  \\
    Fitting procedure               &0.05 \\
    Total                           &0.4  \\
    \hline
    \hline
    \end{tabular}}
\end{table}
\begin{table}[!htbp]
    \centering
    \caption{Systematic uncertainties for the width of g.s. $^6$H.}
    \label{tab2}
    \setlength{\tabcolsep}{4mm}{
    \begin{tabular}{ccc}
    \hline
    \hline
    Uncertainty source   &Uncertainty (MeV) \\
    \hline
    Missing-mass resolution         &0.2  \\
    Longitudinal vertex          &0.05  \\
    Fitting procedure               &0.3 \\
    Total                           &0.4  \\
    \hline
    \hline
    \end{tabular}}
\end{table}

\begin{figure}[!htbp]
  \centering
    \includegraphics[width=0.45\textwidth]{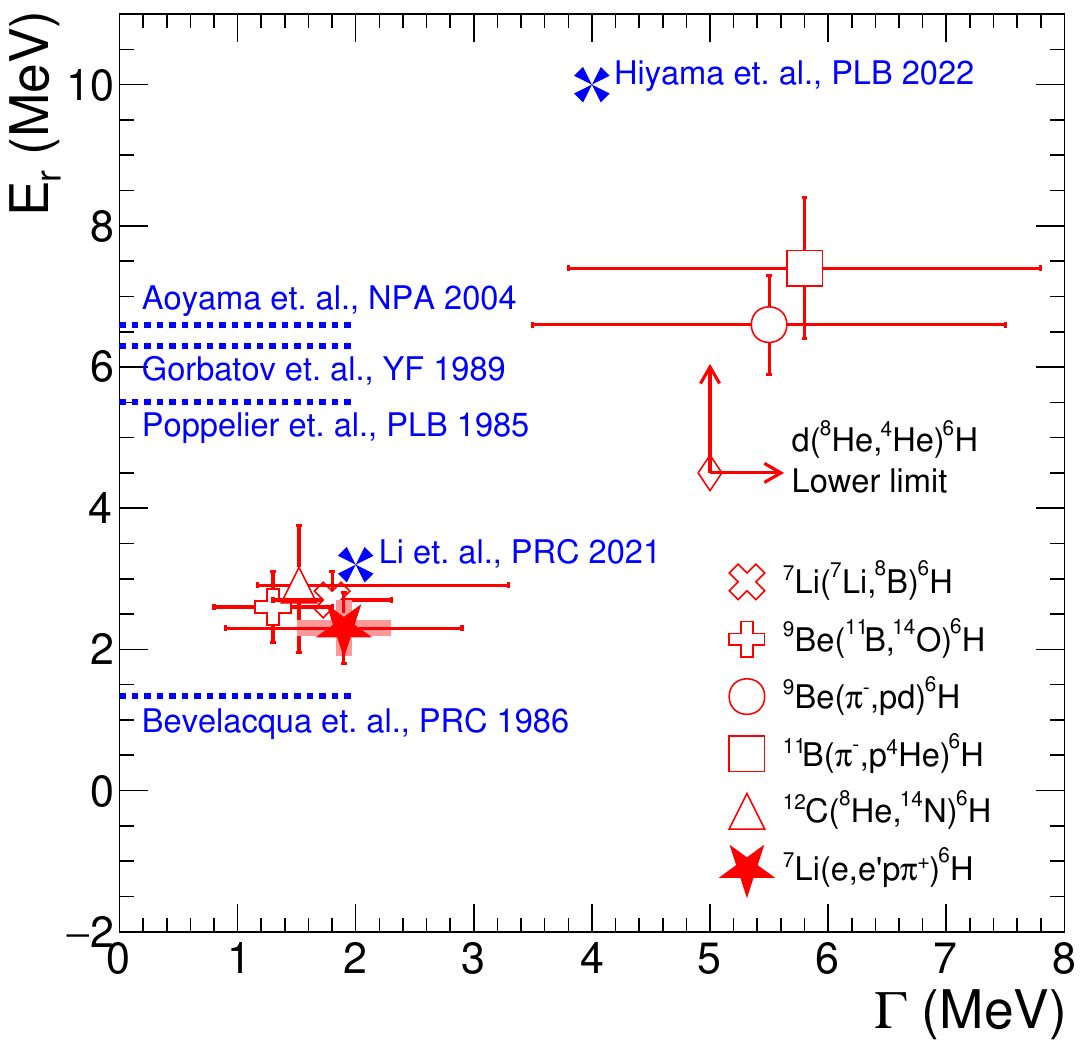}
    \caption{Energies $E_r$ and its corresponding width $\Gamma$ of \hsix ground state. The red star marks the result of this work while red  solid error bar and box representing the statistical and systematic uncertainties respectively. Results from previous experiments \cite{Aleksandrov:1984tb,Belozyorov:1986skf,Gurov:2003pv,Caamano:2008zz} are represented with red hollow markers and their corresponding total uncertainties. The red arrows show the lower limits given in ref. \cite{Nikolskii:2021kqe}. Dashed horizontal lines represent theoretical calculations of \hsix g.s.~energy without width results~\cite{Poppelier:1985vpj,Bevelacqua:1986zz,Gorbatov:1989,Aoyama:2004tay}. Calculations with predicted widths are marked with blue solid markers \cite{Li:2021tyy,Hiyama:2022gzv}.}\label{fig4}
\end{figure}

$Result~and~discussion-$The measured g.s.~energy and corresponding width of \hsix in this experiment are:
\begin{align}
    E_{r}~=~2.3~\pm~0.5~({\rm stat.})~\pm~0.4~({\rm syst.})~{\rm MeV}, \nonumber \\
    \Gamma~=~1.9~\pm~1.0~({\rm stat.})~\pm~0.4~({\rm syst.})~{\rm MeV}. \nonumber
\end{align}
Figure~\ref{fig4} shows our result in comparison with previous experimental measurements and theoretical calculations. Our result favors the measurements giving a small \hsix g.s.~energy of about 2.7~MeV with a small width of about 1.8~MeV. Compared with early data, our result shows a clearer peak structure and a smoother background shape than refs. \cite{Aleksandrov:1984tb,Belozyorov:1986skf} and a much larger statistic than ref. \cite{Caamano:2008zz}. By using the method of particle data group \cite{ParticleDataGroup:2024cfk}, the average values of our results and other three data points at energy close to 2.7~MeV are $E_r=2.6\pm0.3$~(total)~MeV and $\Gamma=1.6\pm0.3$~(total)~MeV. In the experiment measuring a large energy of $\rm ^{6}H$ \cite{Gurov:2003pv}, their measured $\rm ^{5}H$ g.s.~energy at 5.5~MeV with a width at 5.4~MeV \cite{Gurov:2003pv,Gurov:2009} is also larger than the results from other experiments at about 1.8~MeV with a width at about 1.9~MeV \cite{Korsheninnikov:2001buf,Sidorchuk:2003fwa,Golovkov:2005cm}. This may indicate an systematic effect from the type of the reaction. The experimental result of ref.~\cite{Nikolskii:2021kqe} giving a lower limit for \hsix g.s.~energy at about 4.5~MeV excluded the states lower than 3~MeV. Their signal structure was indistinct and the statistic was low from triple coincidence. They also mentioned that if \hsix has a small g.s.~energy, the neutron pairing energy will be absent in the even-neutron nucleus $\rm ^{7}H$ since $\rm ^{7}H$ in their experiment \cite{Muzalevskii:2020svp} has a similar g.s.~energy ($E_{r}\approx2.2$~MeV, with respect to $\rm ^{3}H+4n$) as $\rm ^{6}H$ in our measurement. Our result suggests that this assumption is problematic. 

As shown in Fig.~\ref{fig4} most of the theoretical calculations shown in blue favor a large \hsix g.s.~energy significantly above our result. The authors of ref.~\cite{Li:2021tyy} calculated the g.s.~energy of $\rm ^{4-7} H$ with the Gamow shell model. By reproducing the binding energy of $\rm ^{6-8}He$, their results of $\rm ^{6}H$, whose $E_{r}\approx3.2$~MeV and $\Gamma\approx2$~MeV, are close to our measurements both on the g.s.~energy and its width. Their $\rm ^{7}H$ g.s.~energy ($E_{r}\approx2.4$~MeV) is a little smaller than \hsix but still close to it. However the most recent calculation from ref.~\cite{Hiyama:2022gzv} predicts rather similar but very large g.s. energies of $^6$H and $^7$H around 10~MeV with large widths around 4~MeV by ab initio~\cite{Lazauskas:2019cxj} with an effective n-$\rm ^{3}H$ potential. Our result suggests that stronger n-n or n-n-n interaction may exist in $\rm ^{6}H$ than expect. Our measurement and the latest $^7$H experiment~\cite{Bezbakh:2019dvh} suggest similar g.s. energies of \hsix and $^7$H, though significantly smaller than ref.~\cite{Hiyama:2022gzv}. This phenomenon deviates from the nuclei having similar structures such as $\rm ^{7}He$ ($E\approx-0.6$~MeV, with respect to $\rm ^{4}He+3n$) and $\rm ^{8}He$ ($E\approx-3.2$~MeV, with respect to $\rm ^{4}He+4n$). It suggests that theoretical studies on multi-nucleon interactions need to be reconsidered to understand this experimental result. And more experimental efforts are need to measure the energy of $\rm ^{7}H$ precisely. Indeed, within the framework of our experiment setup, $\rm ^{7}H$ could be produced with a double $\pi^{+}$ producing reaction $\rm ^{7}Li(e, e'\pi^{+}\pi^{+})^{7}H$. However a higher beam intensity will be necessary. Also other light neutron-rich nuclei can be studied with similar electron scattering reactions such as $\rm ^{6}Li(e, e'p\pi^{+})^{5}H$, $\rm ^{4}He(e, e'p\pi^{+})3n$, and $\rm ^{4}He(e, e'\pi^{+}\pi^{+})4n$. Light hypernuclei can be also studied in an electron scattering experiment with Lithium target~\cite{Eckert:2022srr,Chen:2023mel,STAR:2010gyg}.

$Conclusion-$To clarify the puzzling situation of the neutron-rich hydrogen isotope $\rm ^{6}H$, it has been produced for the first time in an electron scattering experiment with the reaction $\rm ^{7}Li(e,~e'p\pi^{+})^{6}H$ at MAMI-A1. Our measurement shows a g.s. \hsix slightly above the $^3$H+n+n+n threshold which suggests a stronger interaction of neutrons in \hsix than expected from the most recent work. Our work may inspire future efforts to study light neutron-rich nuclei with electron scattering reactions.

Dr. Pogge von Strandmann is thanked for measuring the purity of our enriched $\rm ^7Li$ sample. The technical staff and accelerator group are thanked for their support during the experiment. The authors acknowledge the support by the National Key Research and Development Program of China under Grant No. 2022YFA1604900, by the Deutsche Forschungsgemeinschaft (DFG), Germany, through the Research Grant PO 256/8-1. This project has received funding from the European Union’s Horizon 2020 research and innovation programme under Grant No. 824093. The support from National Natural Science Foundation of China under Grant Nos. 12025501 and 12147101, from JSPS, Japan KAKENHI Grant No. 24H00219 are also thanked.

\bibliography{refv0}
\newpage
\section{Appendix}
\begin{figure}[!htbp]
  \centering
    \includegraphics[width=0.45\textwidth]{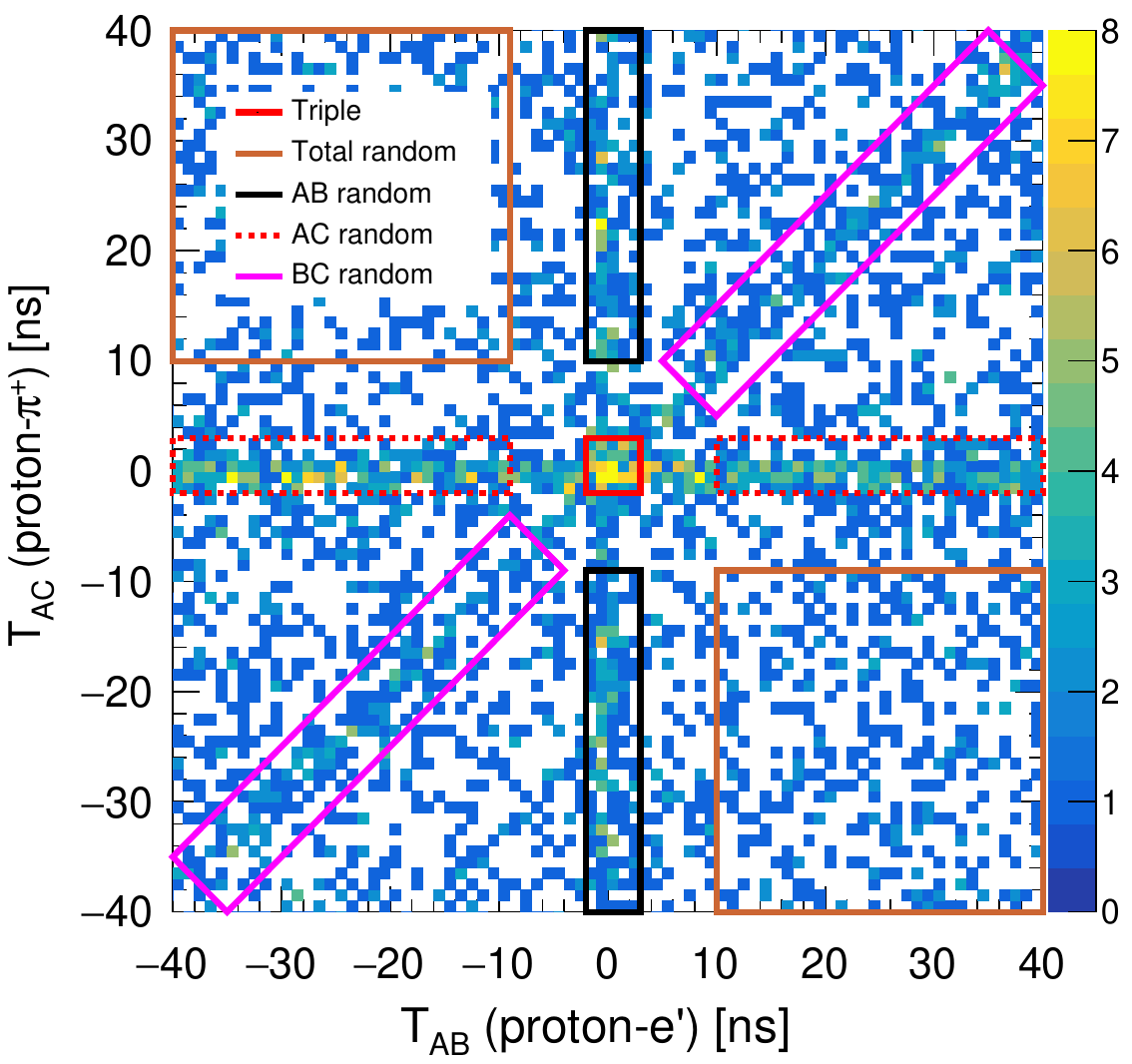}
    \caption{The 2 dimensional plot of coincident times between different spectrometer pairs. The x axis is the coincidence time between A and B. The y axis is the coincidence time between A and C. The region enclosed with red square represents the triple coincidence events for the signal. Regions marked with black, dashed red, and magenta rectangles represent events from random double coincidence. Events in the brown squares come from total random coincidence.}\label{fig2}
\end{figure}

$Triple~coincidence-$Figure~\ref{fig2} shows the two dimensional plot of double coincidences of the spectrometers. The x axis represents the coincidence time between protons and electrons from the timing systems of spectrometers A and B. The y axis is the coincidence time between protons in spectrometer A and $\pi^{+}$ in C. Naturally, the diagonal line of $T_{\rm AB}$ and $T_{\rm AC}$ shows up which represents coincidences between electrons and $\pi^{+}$ from B and C. The events in the intersection region of them, shown in the red square in the center with both 5~ns width timing windows for $T_{\rm AB}$, $T_{\rm AC}$, and $T_{\rm BC}$, are selected as the triple coincidence events. With the constraints of specific energy loss and time of flight, the purities of three particles are practically 100\%. To evaluate the background caused by the random coincidences, the events from other double coincidence regions beyond the triple region, shown as the black, dashed red, and magenta rectangles, and the constant background of total random events, which are marked by the brown regions, were properly weighted with respect to the area of the triple coincidence region in the center of the Fig.~\ref{fig2} and summed up (see ref. \cite{Makek:2016utb} for details).

\end{document}